
\magnification=1200
\baselineskip=19pt
\overfullrule=0pt

\def\rt{railroad trestle }
\def\sa{sawtooth }

\centerline{\bf On Quantum Solitons in the Sawtooth Lattice}
\vskip .4in

\centerline{Diptiman Sen and B. Sriram Shastry,}

\centerline{\it Indian Institute of Science, Bangalore 560012, INDIA}

\centerline{R. E. Walstedt and R. Cava,}

\centerline{\it AT\& T Bell Laboratories, Murray Hill, New Jersey 07974, USA}
\vskip .2in

\line{\bf ABSTRACT \hfill}

We study the sawtooth lattice of a coupled spin $1/2$ Heisenberg system,
a variant of the railroad trestle lattice. The ground state of this system
is two-fold degenerate with periodic boundary conditions and supports kink
antikink excitations, which are distinct in this case, unlike the railroad
trestle lattice. The resulting low temperature thermodynamics is compared
with the recently discovered Delafossites $Y Cu O_{2.5} ~$.

\vfill
\eject

\line{\bf I. ~ INTRODUCTION \hfill}
\vskip .3in

Several novel quantum toplogical excitations have been
explored in the quantum antiferromagnetic
spin systems in the last decade, expanding on the
Anderson-Kubo spin waves. These include ``spinons'',
i.e. spin $1/2$ objects of Faddeev and Takhtajan [1],
a name coined by Anderson, describing the excitations of
the isotropic one-dimensional Heisenberg spin $1/2$ antiferromagnet (AFM),
with a concomitant two-parameter elementary excitation spectrum with
an asymptotic four-fold degeneracy. The anisotropic Heisenberg AFM
in the Neel ordered, massive phase contains ``domain walls''
found by Johnson, Krinsky and McCoy [2] from Baxter's solution
of the XYZ model. These domain walls
separate two regions of Ising like ordered states, and propagate
as dressed fermions [3] (again these exhibit a four-fold degeneracy).
The domain walls would broaden out and indeed the
width would diverge with the correlation length as the isotropic
point is reached, so the limiting
excitation would have a delocalized character.
It may be tempting to view the ``spinons'' as limiting
cases of the domain walls, although the AFM order vanishes at the
isotropic point.  Yet another class of excitations were introduced
by Shastry and Sutherland [SS] [4] in the context of models with broken
translation symmetry [5]. These are topological quantum
solitons separating two regions of broken translational symmetry and have
again a four-fold degeneracy of two-parameter excitations. In addition,
these are fairly compact objects with a width
consisting of a few lattice constants. In view of
the theoretical interest in these constructs, it is
interesting to ask for experimental realizations of such systems.

Recently a new family of Delafossite compounds have been synthesized
which seem to be promising from this viewpoint.
The $Y Cu O_2 ~$ lattice consists of planes of coupled $Y O_2 ~$
octahedra linked by two-fold-coordinated bridging $Cu^+ ~$ ions, which
form a triangular planar array. It is possible to intercalate $O^{2-} ~$
ions into the $Cu^{+} ~$ planes, forming different lattice symmetries
depending on the amount of intercalant. For compositions up to and
including $Y Cu O_{2.5} ~$, the planar $O^{2-} ~$ form an orthorhombic
structure [6]. At the upper limit of this range one has a magnetic
insulator, with all of the copper ions converted to $Cu^{2+} ~$.
For compositions with additional oxygen beyond $O_{2.5} ~$, the
structural symmetry becomes hexagonal and mobile hole carriers appear,
rendering the $Cu O_x ~$ planes weakly conducting [7].

Preliminary data on the structure of the orthorhombic phase for
$Y Cu O_{2.5} ~$ suggests two sets of locally
triangular configuration of copper sites with $O^{2-} ~$ ions in the
center of one of the sets of triangles, providing superexchange paths (Fig.1).
 The two sets of triangles are separated and have no $O{2-} ~$ ions in their
midst, so we have a good possibility of one-dimensional exchange
coupled $Cu^{2+} ~$ i.e. a $S=1/2$ system. We have carried out NMR
measurements of the relaxation rates $1/T_1, \; 1/T_2 ~$
between $70^o K$ and $230^o K$. It is evident from the NMR results that
there are substantial exchange couplings between the $S=1/2$ $Cu^{2+} ~$
spin moments in this structure.
We find an activated behavior as a function of $T$ with a gap
$\sim 650^o K$. Such couplings would arise between nearest
neighbor spins to an $O^{2-} ~$ ion, which would act as the conduit
for $120^o ~$ exchange paths. If we presume that such couplings
between nearest neighbor and second-neighbor spins to an $O^{2-} ~$ ion are
negligible ($60^o ~$ exchange paths), then the system divides into
a series of parallel one-dimensional "sawtooth" lattices of exchange
coupled spins, which are only weakly interacting. Such a system,
the sawtooth lattice is analyzed in this paper, and shown to have an
interesting and unique set of
magnetic excitations, namely quantum solitons very similar to the ones
discussed in SS, with a new feature, namely the kink antikink symmetry in the
Majumdar model is broken here.

\vskip .3in
\line{\bf II. ~THE SAWTOOTH CHAIN: ANALYTICAL RESULTS \hfill}
\vskip .3in

The Hamiltonian for the \sa chain may be written as a sum of Hamiltonians
governing triangles of spins (see Fig. 2)
$$\eqalign{H ~&=~ \sum_{n=1}^N ~H_n \cr
H_n ~&=~ {J \over 2}~ \Bigl[ ~(~{\vec S}_{2n-1} ~+~ {\vec S}_{2n} ~+~
{\vec S}_{2n+1} ~)^2 ~-~ {3 \over 4} ~\bigr] ~. \cr}
\eqno(2.1)$$
Here $N$ is the number of triangles and $J$ denotes the antiferromagnetic
coupling. We may consider either open chains (with $2N+1$ sites) or
periodic chains (with $2N$ sites).

To find the ground states of (2.1), we note that $H_n ~$ is proportional
to a projection operator since its eigenvalues are $0$ (if the total spin
in triangle $n$ is $1/2$) and $3J/2$ (if the total spin is $3/2$). For two
sites $i$ and $j$ with $i < j$, we denote the singlet state mathematically
by $[i,j]=(\vert \alpha_i \beta_j \rangle - \vert \beta_i \alpha_j
\rangle )/ {\sqrt 2}$, and pictorially by a double line joining $i$ and $j$
as indicated in Figs. 3 and 4. (Here $\alpha_i ~$ and $\beta_i ~$ denote spin
up and down respectively at site $i$).
Then the states with total spin $1/2$ in triangle $n$ can be thought of as
either $[2n-1,2n]$ with the spin at site $2n+1$ free (either $\alpha$ or
$\beta$), or as $[2n,2n+1]$ with the spin at $2n-1$ free. The other possible
pairing $[2n-1,2n+1]$ is linearly dependent since
$$[2n,2n+1] ~\alpha_{2n} ~=~ [2n-1,2n] ~\alpha_{2n+1} ~+~ [2n,2n+1] ~
\alpha_{2n-1}
\eqno(2.2)$$
It is now easy to show that the periodic chain has two degenerate ground
states (with energy $E_o ~=~ 0$) given by
$$\eqalign{\vert ~I~ \rangle ~&=~ \prod_{n=1}^N ~[2n-1,2n] \cr
{\rm and} \quad \vert ~II~ \rangle ~&=~ \prod_{n=1}^N ~[2n,2n+1] ~, \cr}
\eqno(2.3)$$
where ${\vec S}_{2N+1} ~\equiv ~{\vec S}_1 ~$ [4, 8]. The open chain has
$2(N+1)$degenerate ground states (with $E_o ~=~ 0$) given by
$$\eqalign{\vert ~2n+1, ~\alpha ~{\rm or}~ \beta ~\rangle ~=~ &\bigl( ~
\prod_{n=1}^m ~ [2n-1,2n] ~\bigr) ~\bigl( ~\prod_{n=m+1}^N ~[2n,2n+1] ~
\bigr) ~~{\rm times} \cr
& \alpha_{2m+1} ~~{\rm or} ~~\beta_{2m+1} ~, \cr}
\eqno(2.4)$$
where $m$ may take any value from $0$ to $N$. Such a state is shown in
Fig. 3. This configuration can be thought of as a kink at site $2m+1$
which separates the ground state I on its left from the ground state
II on its right. Since all states in (2.4) have the same energy (namely,
zero), a linear combination of them like
$$\vert ~k ~\rangle ~=~ {1 \over {\sqrt N}} ~\sum_{m=0}^N ~\exp (ikm) ~
\vert ~ 2m+1, \alpha ~\rangle
\eqno(2.5)$$
has the same energy for all values of $k$. In the limit $N \rightarrow
\infty$, (2.5) denotes a momentum eigenstate. We therefore see that kinks
in the \sa chain have the dispersionless spectrum $\omega (k) =0$. Further,
kinks only differ from ground states I and II at a single site.

We will see below that {\it antikinks} are quite different in that they
have a non-trivial dispersion and they do not just differ from the states
I and II at only one site. In fact, we are unable to solve for the wave
function and dispersion of antikinks exactly. The best we can do is to
variationally estimate these quantities more and more accurately by
considering antikink configurations spread over $1$ site, $5$ sites and
so on.

We would like to make a few comments before examining the antikinks.
Firstly, it can be shown by induction that the states in (2.3 - 2.4) are
indeed the only ground states for the \sa chain [9]. Secondly, it can
be proved rigorously that there is a finite gap between the ground states
and the first excited state [10, 11]. Our discussion of antikinks will lead
to an accurate estimate of this gap. Finally, the situation here may be
contrasted with that obtaining in the \rt which was first studied by
Majumdar and coworkers [5]. The Hamiltonian for the \rt differs from (2.1)
in also having a $H_n ~$ for the sites $(2n,2n+1,2n+2)$. As a result, this
model only has the two ground states of types I and II (except possibly
for free spins at the ends if the chain is open). There is a finite gap
to excited states. Kinks and antikinks are on the same footing in the
railroad trestle. They are not exactly solvable but they can be shown to
have identical dispersions.

We now study antikinks to a first approximaton by considering a state
like the one shown in Fig. 4 (a). This is a configuration which has ground
state II to the left of a $1$-site cluster (located at site $2n$) and
ground state I to its right. We denote this state by $\vert ~2n ~\rangle_1 ~$.
(For simplicity of notation, we will henceforth drop the spin index, $\alpha$
or $\beta$, of the free spin). In the limit $N \rightarrow \infty$, we
consider a momentum eigenstate
$$\vert ~ k ~\rangle ~=~ {1 \over {\sqrt N}} ~\sum_n ~\exp (ikn) ~\vert ~
2n ~\rangle_1 ~.
\eqno(2.6)$$
This state has a non-trivial norm because
$${}_1 \langle ~ 2n ~\vert ~2m ~ \rangle_1 ~=~ (-1)^{n-m}~ / ~2^{\vert
n-m \vert} ~.
\eqno(2.7)$$
Hence
$$\langle ~k ~\vert ~k ~\rangle ~=~ 3 / (5 + 4 \cos k ) ~.
\eqno(2.8)$$
Further
$${}_1 \langle ~2n ~ \vert ~H_l ~\vert ~ 2m ~\rangle_1 ~=~ {3 \over 4} ~J ~
\delta_{nl} ~ \delta_{lm}
\eqno(2.9)$$
implies that
$$\langle ~k ~\vert ~H~ \vert ~ k~\rangle ~=~ {3 \over 4} ~J~.
\eqno(2.10)$$
Our estimate of the dispersion based on this $1$-cluster calculation is
therefore
$$\omega (k) ~=~ ( {5 \over 4} + \cos k) ~J ~.
\eqno(2.11)$$
(The kinks and antikinks in the \rt have the same dispersion as in (2.11)
for the $1$-cluster approximation). Eq. (2.11) has a minimum at $k=\pi$
where the antikink energy is $J/4$. This is our first estimate of the gap
in the open chain and, as we will argue below, in the periodic chain also.

We may now improve our estimate by considering $3$-cluster configurations.
Due to Eq. (2.2), however, a cluster of three neighboring sites with
$S=1/2$ can be reduced to a superposition of $1$-cluster states like
$\vert ~ 2n~\rangle_1 ~$. So we have to continue on to $5$-cluster
configurations. The only two linearly independent configurations that we
need to consider are the ones
shown in Figs. 4 (b) and (c). We denote these two by $\vert ~ 2n ~\rangle_2 ~$
and $\vert ~ 2n ~\rangle_3 ~$ respectively where $2n$ denotes the center of the
$5$-clusters. Note that $\vert ~2n ~\rangle_2 ~$ and $\vert ~2n~ \rangle_3 ~$
are related to each other by reflection about the site $2n$. We now consider a
momentum eigenstate with two complex variational parameters $a$ and $b$
$$\vert ~k, a, b~\rangle ~=~ {1 \over {\sqrt N}} ~\sum_n ~\exp (ikn) ~\Bigl[ ~
\vert ~2n ~\rangle_1 ~+~ a~ \vert ~2n~\rangle_2 ~+~ b~ \vert ~2n~\rangle_3 ~
\Bigr] ~,
\eqno(2.12)$$
and minimize $\omega (k,a,b)$ by varying $a$ and $b$. The computation
is straightforward though lengthy. We first obtain the overlaps
$$\eqalign{{}_2 \langle 2n ~\vert ~2m~ \rangle_1 ~&=~ - (-1)^{n-m} ~/ ~
2^{\vert n-m \vert} \quad {\rm if} \quad n \ge m+1 ~, \cr
&=~ - 1 / 4 \quad {\rm if} \quad n = m ~, \cr
&=~ {1 \over 2} ~(-1)^{n-m} ~/ ~2^{\vert n-m \vert} \quad {\rm if} \quad
n \le m-1 , \cr}
\eqno(2.13)$$
$$\eqalign{{}_2 \langle 2n ~\vert ~2m~ \rangle_2 ~&=~ 1 \quad {\rm if} \quad
n = m ~, \cr
&=~ - 1 / 8 \quad {\rm if} \quad \vert n-m \vert =1 ~, \cr
&=~ - {1 \over 2} ~(-1)^{n-m} ~/ ~2^{\vert n-m \vert} \quad {\rm if} \quad
\vert n-m \vert \ge 2 , \cr}
\eqno(2.14)$$
$$\eqalign{{}_2 \langle 2n ~\vert ~2m~ \rangle_3 ~&=~ (-1)^{n-m} ~/ ~
2^{\vert n-m \vert} \quad {\rm if} \quad n \ge m+2 ~, \cr
&=~ - 1 / 8 \quad {\rm if} \quad n = m+1 ~, \cr
&=~ - 1 / 2 \quad {\rm if} \quad n = m ~, \cr
&=~ 1 / 4 \quad {\rm if} \quad n = m-1 ~, \cr
&=~ {1 \over 4} ~(-1)^{n-m} ~/ ~2^{\vert n-m \vert} \quad {\rm if} \quad
n \le m-2 . \cr}
\eqno(2.15)$$
All other overlaps can be obtained by using the reflection symmetry mentioned
above. Thus,
$$\eqalign{{}_3 \langle ~2n ~\vert ~2m~ \rangle_1 ~&=~{}_2 \langle ~2m ~
\vert ~2n~ \rangle_1 , \cr
{\rm and} \quad {}_3 \langle ~2n ~\vert ~2m~ \rangle_3 ~&=~ {}_2 \langle ~
2n ~ \vert ~2m~ \rangle_2 , \cr}
\eqno(2.16)$$
We therefore obtain
$$\eqalign{\langle ~k~ \vert ~k~ \rangle ~=~& {1 \over {5 + 4 \cos k}}~
\Bigl[ ~ (5+4 \cos
k)~ A_1 ~+~ (4 + 2 e^{-ik} ~)~ A_2 ~+~ (4 +  2 e^{ik} ~)~ A_2^\star ~\Bigr] ~,
\cr
{\rm where} \quad A_1 ~=~& 1 ~-~ {1 \over 4} (a+a^\star +b+b^\star ) ~+~{1
\over 4} (aa^\star + bb^\star )~(4- \cos k) \cr
&- ~{1 \over 2} (ab^\star +ba^\star) ~+~ {1 \over 8} ab^\star (2 e^{-ik} ~
-~ e^{ik} ~)~+~ {1 \over 8} ba^\star (2 e^{ik} ~-~e^{-ik} ~) , \cr
{\rm and} \quad A_2 ~=~ &{{e^{ik}} \over 4} ~(-2-a^\star +2a -b +2b^\star)~
+~ {{e^{i2k}} \over 16} ~(-2aa^\star -2bb^\star +4ab^\star +ba^\star) ~.\cr}
\eqno(2.17)$$
Next we calculate the matrix elements of $H_n ~$. Thus
$$\eqalign{{}_2 \langle 2n-2 ~\vert ~H_n ~\vert ~2n ~\rangle_1 ~&=~ - ~{3
\over 8} ~J \cr
{}_2 \langle 2n+2 ~\vert ~H_n ~\vert ~2n ~\rangle_1 ~&=~  {3 \over 8} ~J \cr
{}_2 \langle 2n-4 ~\vert ~H_{n-1} ~\vert ~2n ~\rangle_2 ~&=~-~ {3 \over
{16}} ~J \cr
{}_2 \langle 2n ~\vert ~H_{n-1} ~\vert ~2n ~\rangle_2 ~&=~ {3 \over 4} ~J \cr
{}_3 \langle 2n-4 ~\vert ~H_{n-1} ~\vert ~2n ~\rangle_2 ~&=~ {3 \over
{16}} ~J \cr
{}_3 \langle 2n ~\vert ~H_{n-1} ~\vert ~2n ~\rangle_2 ~&=~-~ {3 \over 8} ~
J \cr
{}_2 \langle 2n-4 ~\vert ~H_{n-1} ~\vert ~2n ~\rangle_3 ~&=~ {3 \over
{16}} ~J ~. \cr}
\eqno(2.18)$$
All other matrix elements can either be obtained from the above by
translation or reflection symmetry, or are zero. We then find that
$$\eqalign{\langle ~k~ \vert ~H~ \vert ~k~\rangle ~=~ &{3 \over 4} ~+~ i ~{3
\over 4} ~\sin k ~ (a-a^\star ~-b+b^\star ~) \cr
&+ ~({3 \over 2} - {3 \over 8} \cos 2k) (aa^\star +bb^\star ~) ~+~ ({3 \over
8} \cos 2k - {3 \over 4} ) (ab^\star ~+~ ba^\star ~) ~.\cr}
\eqno(2.19)$$

We have found that the minimum value of $\omega (k)$ occurs at $k= \pi$ if
we take $a=b$ to be real. Then
$$\omega (\pi; a) ~=~ {1 \over 4} ~{{1+2a^2} \over {1-a+a^2 /2}} ~.
\eqno(2.20)$$
This has a minimum at $a=(3-{\sqrt {17}})/4 = - 0.2808 $ where $\omega =
0.2192 ~J$. (For the railroad trestle, $\langle k \vert k \rangle$ is the
same as in (2.17) while $\langle k \vert H \vert k\rangle$ has the extra
term $3(aa^\star ~+bb^\star ~)/4$ on the right hand side of (2.19). Hence
$$\omega (\pi, a) ~=~ {1 \over 4} ~{{1+4a^2} \over {1-a+a^2 /2}} ~.
\eqno(2.21)$$
whose minimum value is $0.2344 ~J~$).

We see that the estimate of the gap changes relatively little on going from
$1$-clusters to $5$-clusters. This is because of the small correlation
length $\xi$ in this system. We expect that the estimate of the gap
$\Delta (l)$ from an $l$-cluster calculation will differ from the true gap
$\Delta (\infty)$ by terms of order $\exp (-l / \xi )$. The gap in the
\rt chain has been estimated from a $9$-cluster calculation in Ref. 12. From
the values $\Delta (1) = 0.25 J$, $\Delta (5) = 0.2344 J$ and $\Delta (9) =
0.2340 J$, we estimate that $\xi \sim 1.1$. While we have not computed $\Delta
(9)$ for the \sa chain, we expect that it will differ very little from $\Delta
(5)$ for a similar reason.

To summarize so far, we have seen that kinks have zero dispersion while
antikinks have a dispersion with a gap of $0.2192 ~J$ at $k= \pi$. We now
identify the latter figure with the gap in the open chain. This assumes that
there are no bound states of several kinks and antikinks which have a lower
energy. For the railroad trestle, it is in fact known that there is no bound
state of a kink and an antikink which has lower energy than a well-separated
kink and antikink [4].

We will now argue that a periodic chain has the same gap and, further, that it
has a dispersionless spectrum for its lowest excitation. Any excitation in a
periodic chain must consist of a succession of alternating kinks and antikinks.
In the absence of low-energy bound states, the lowest excitation in a long
periodic chain will consist of one kink well-separated from one antikink. The
energy of this state is the sum
$$\omega (Q) ~=~ \omega_K (k_1 ) ~+~ \omega_{\overline K} (k_2 ) ~,
\eqno(2.22)$$
where $k_1 $ and $k_2 $ denote the momenta of the kink ($K$) and the
antikink (${\overline K}$), and the total momentum of this state is
$Q = k_1 + k_2 ~$. It is now clear that since $\omega_K (k_1 )= 0$ for all
$k_1 $, the minimum possible value of $\omega (Q)$ is given by the constant
$\Delta \equiv \omega_{\overline K} ~(\pi) $ since we can always choose
$k_1 = Q- \pi $.

Indeed, numerical studies of finite periodic chains upto $N=10$ by Kubo had
indicated the existence of a dispersionless spectrum with $\omega (Q)
\simeq 0.219 ~J$ for all $Q$ [13]. We now have the explanation of this
striking property of the periodic \sa as arising from the dispersionless
spectrum of the kink. Further, our $5$-cluster computation has already yielded
an estimate of the gap which is very close to the value obtained numerically.

We may now use the above results to study low-temperature thermodynamic
properties of the \sa chain. For instance, we can estimate the magnetic
susceptibility based on the picture of a low density of alternating kinks
and antikinks which are well-separated and noninteracting. In the presence
of an external magnetic field $B$, they have the energies $-2 \mu S_z ~$
and $\omega_{\overline K} (k) - 2 \mu S_z ~$ respectively where $\mu$ is the
Bohr magneton. If we use the $1$-cluster expression for $\omega_{\overline K}
(k)$ given in Eq. (2.11), we obtain the partition functions for one kink
and one antikink as
$$\eqalign{x_K ~&=~ 2 \cosh (\mu \beta B) \cr
{\rm and} \quad x_{\overline K} ~&=~ 2 \cosh (\mu \beta B) ~I_o (\beta J)
\exp (-{5 \over 4} \beta J) \cr}
\eqno(2.23)$$
respectively, where $\beta = 1 / k_B T$ is the inverse temperature and
$I_o ~$ is a modified Bessel function. The free energy per site is then given
by
$$f ~=~ - ~{1 \over \beta} ~(x_K ~x_{\overline K} ~)^{1/2} ~,
\eqno(2.24)$$
and the magnetic susceptibility per site $\chi = -(\partial^2 f ~/~ \partial
B^2 ~)_{B=0} ~$ is
$${{\chi J} \over {2 \mu^2}} ~=~ \beta J ~\Bigl[ ~I_o (\beta J) \exp (-{5
\over 4} \beta J) ~\Bigr]^{1/2} ~.
\eqno(2.25)$$
Note that this thermodynamic quantity exhibits a gap equal to $J/8$ at very
low temperature which is half the sum of the gaps for the kink (zero) and
the antikink ($J/4$). In Fig. 5, we show the magnetic susceptibility as a
function of $k_B T /J$.

Coming back to the system $Y Cu O_{2.5} ~$, we see that the gap of
$\sim 0.22 J$, if equated to the NMR activation energy, implies
that $J \sim 3000 ^) K$, which is rather too large. Indeed, the
largest $J's$ are a factor of $2$ smaller than this, as measured
in the high $T_c ~$ systems, which have comparable $Cu-O$ bond lengths
as in these compounds, namely $\sim 2 A^o ~$. It then seems likely that
these systems either do not allow for a decoupling between these
sawtooth lattices, forcing say a pair of these excitations, or
else there might be pairwise dimerization, which could be signalled in
detailed structural studies. It would be interesting to study these issues
in detail experimentally, as would the ESR on these systems at elevated
temperatures, say $700^o K$, where the four-fold degeneracy would be
reflected in the existence of free spin $1/2$ excitations with a thermally
activated density.

\vfill
\eject

\line{\bf REFERENCES \hfill}
\vskip .3in

\item{1.}{L. Faddeev and L. Takhtajan, Phs. Letts. {\bf 85 A}, 375 (1981).}

\item{2.}{J. D. Johnson, B. McCoy and S. Krinsky, Phys. Rev. A {\bf 8},
2526 (1973).}

\item{3.}{G. Gomez-Santos, Phys. Rev. B {\bf 41}, 6788 (1990).}

\item{4.}{B. S. Shastry and B. Sutherland, Phys. Rev. Lett. {\bf 47}, 964
(1981).}

\item{5.}{C. K. Majumdar, J. Phys. C {\bf 3}, 911 (1969); C. K. Majumdar
and D. K. Ghosh, J. Math. Phys. {\bf 10}, 1388 and 1399 (1969); C. K.
Majumdar, K. Krishan and V. Mubayi, J. Phys. C {\bf 5}, 2896 (1972).}

\item{6.}{ R. J. Cava {\it et. al.}, J. Sol. St. Chem, {\bf 104}, 437 (1993).}

\item{7.}{ R. E. Walstedt {\it et. al}, Phys. Rev. {\bf B 49}, 12369 (1994).}

\item{8.}{M. W. Long and R. Fehrenbacher, J. Phys. Condens. Matter {\bf 2},
2787 (1990).}

\item{9.}{F. Monti and A. S$\ddot u$t$\ddot o$, Phys. Lett. A {\bf 156},
197 (1991).}

\item{10.}{I. Affleck, T. Kennedy, E. H. Lieb and H. Tasaki, Comm. Math.
Phys. {\bf 115}, 477 (1988).}

\item{11.}{S. Knabe, J. Stat. Phys. {\bf 52}, 627 (1988).}

\item{12.}{W. J. Caspers, K. M. Emmett and W. Magnus, J. Phys. A {\bf 17},
2687 (1984).}

\item{13.}{K. Kubo, Phys. Rev. B {\bf 48}, 10552 (1993).}

\vfill
\eject

{\bf FIGURE CAPTIONS \hfill}
\vskip .3in

\item{1.}{A picture of the $Cu O$ planes in $Y Cu O_{2.5} ~$ from
preliminary structural data.}

\item{2.}{The \sa chain. The three sites forming triangle $n$ are numbered
as shown.}

\item{3.}{A kink configuration with the free spin at site $2n+1$. The double
lines indicate singlets.}

\item{4.}{The three antikink configurations centred about site $2n$. (a)
is a $1$-cluster while (b) and (c) are $5$-clusters.}

\item{5.}{Low-temperature magnetic susceptibility as a function of
$k_B T /J$.}

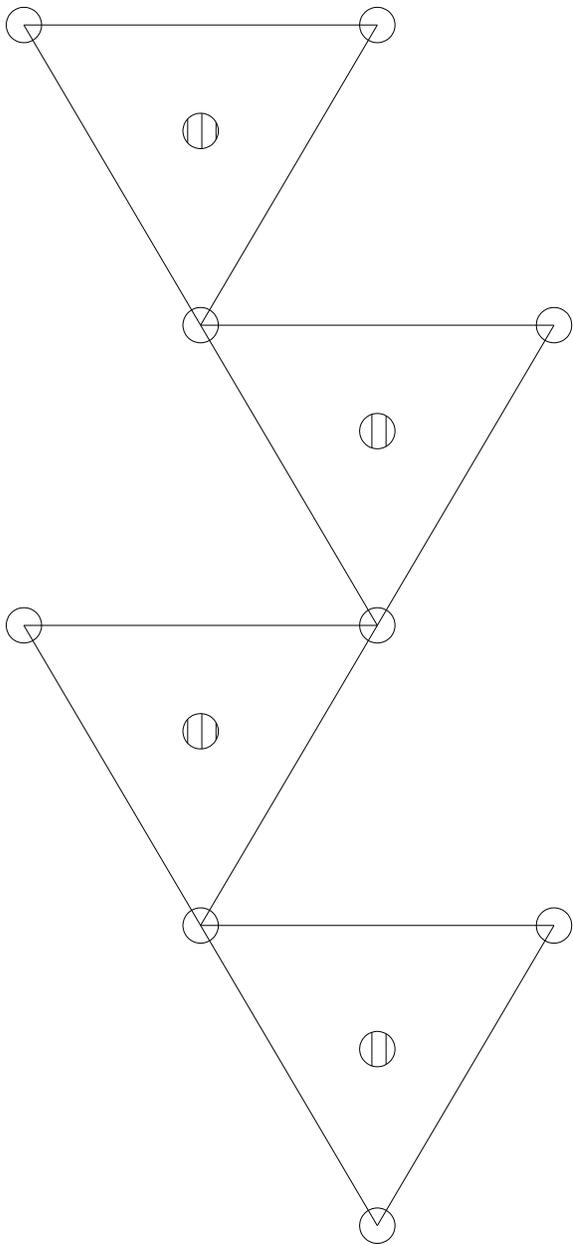
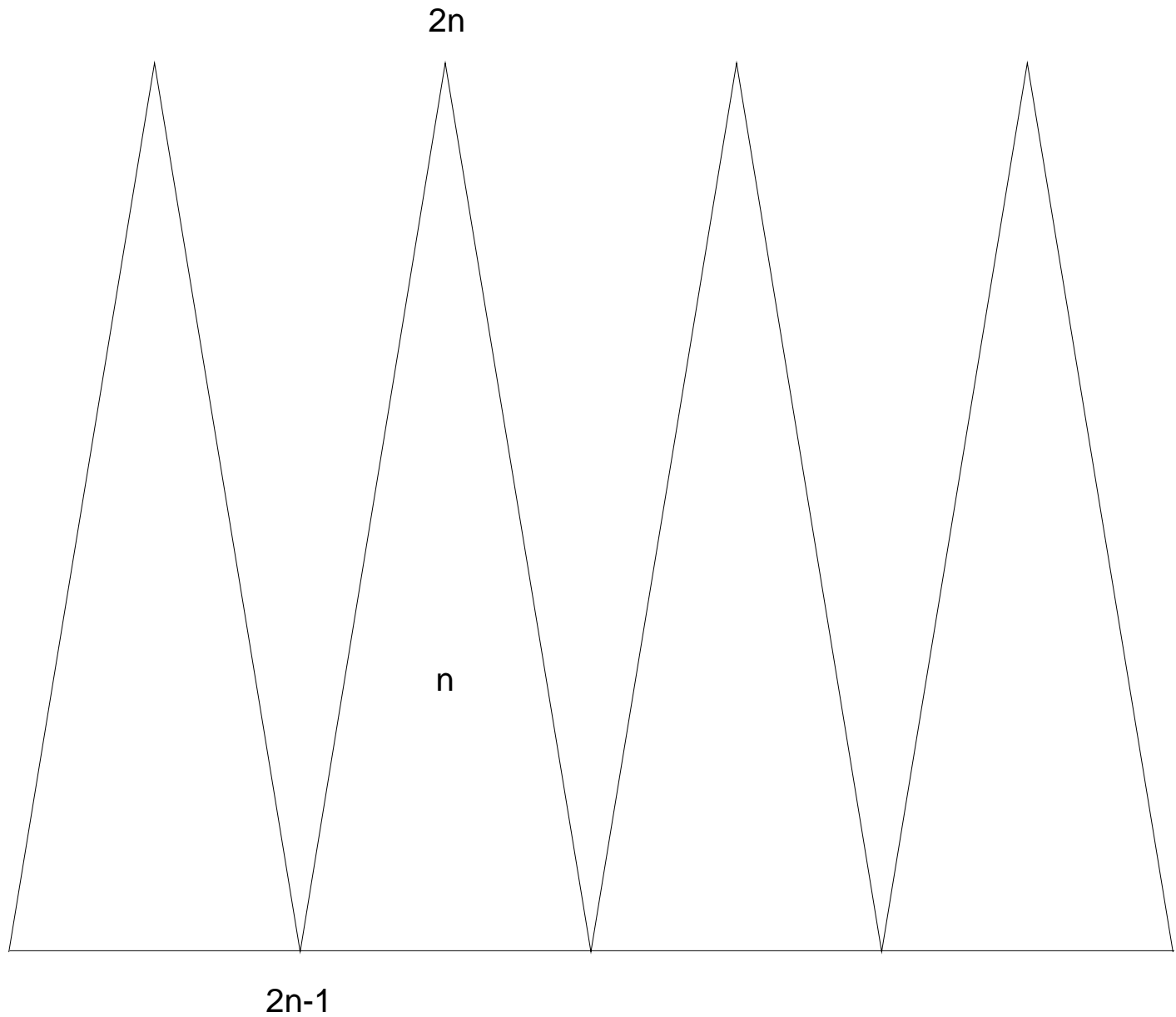
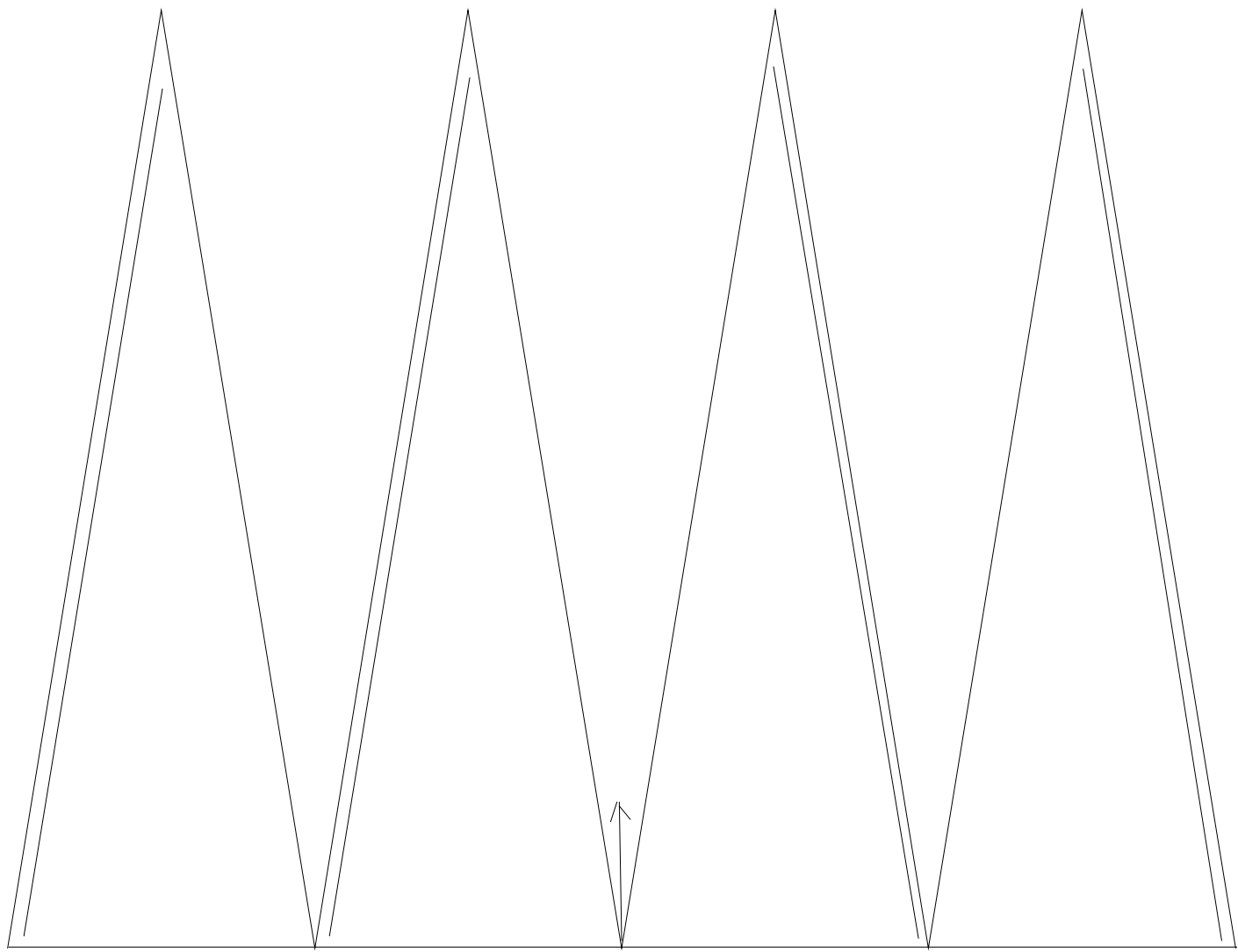
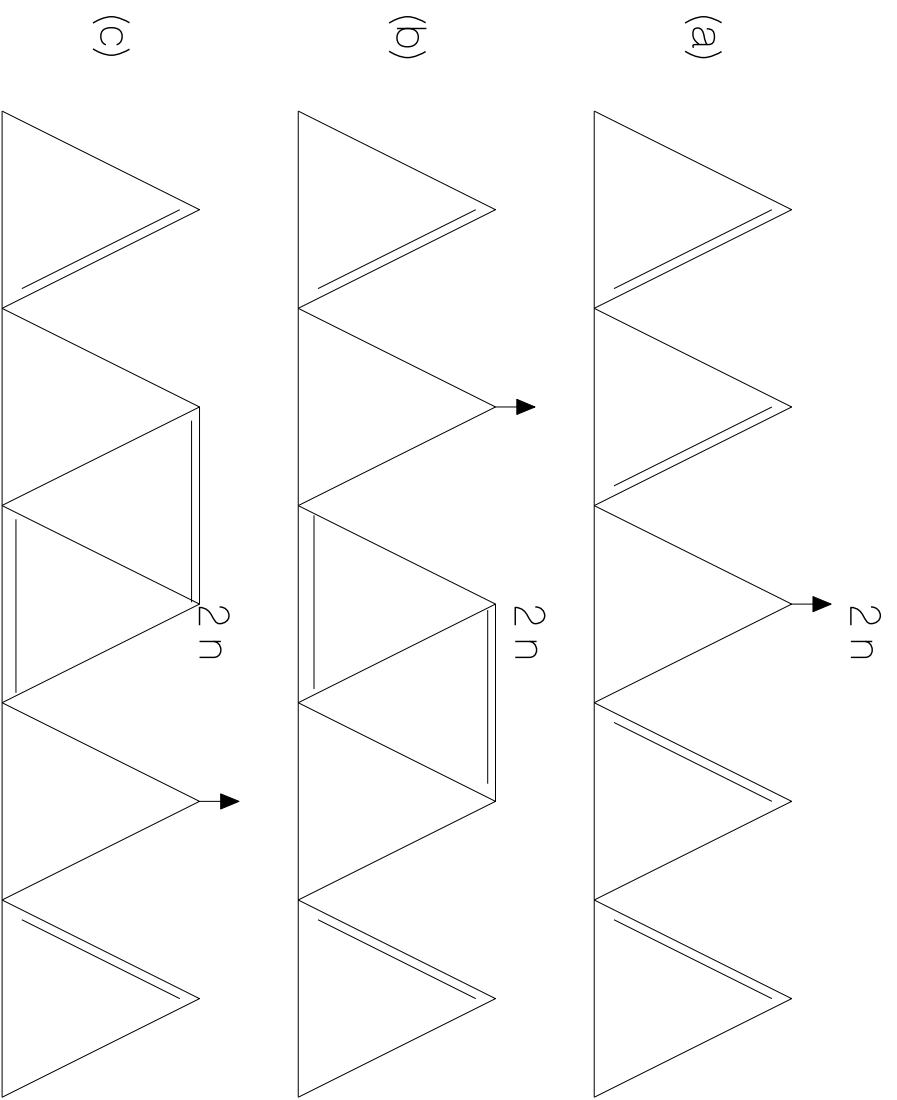
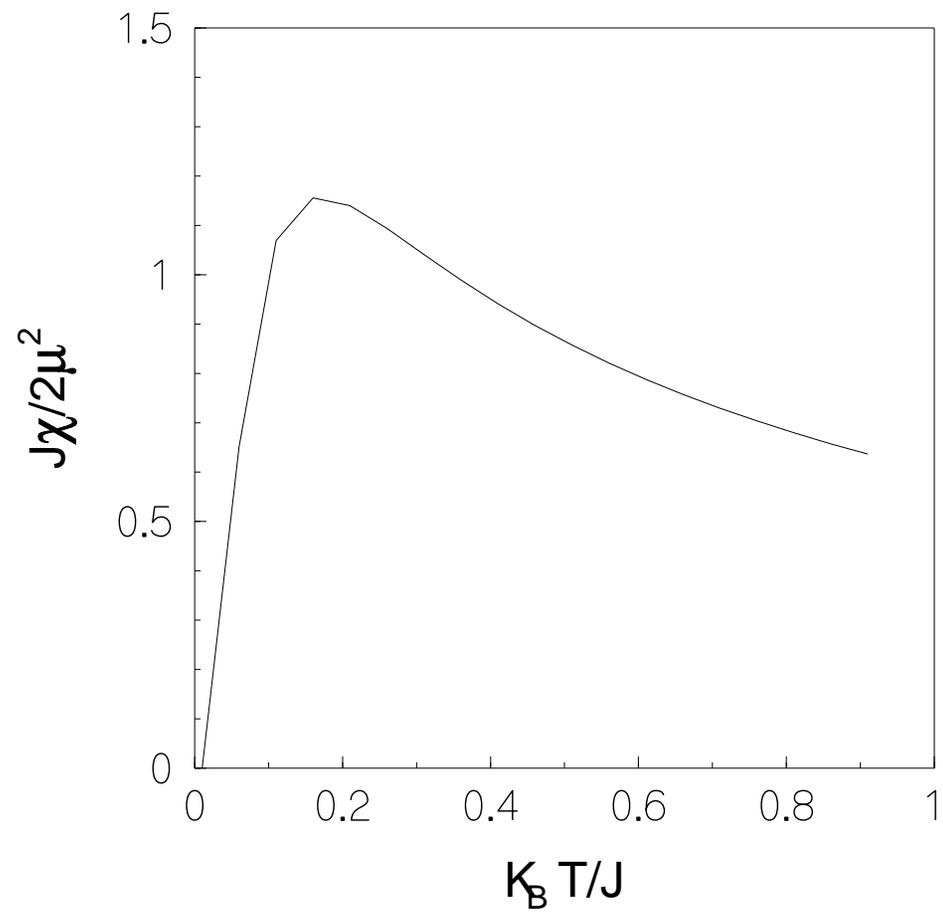
\end